\documentclass[aps,prc,superscriptaddress,twoside,twocolumn,nofootinbib,%
showpacs,dvips]{revtex4}
\usepackage{mathrsfs,bm}
\usepackage{longtable,lscape}
\usepackage{txfonts}
\usepackage{amssymb}
\usepackage{indentfirst}
\usepackage{graphicx,,booktabs}
\usepackage{color}
\usepackage{amssymb}

\def\half{{\textstyle{1\over 2}}}

\def\thalf{{\textstyle{3\over 2}}}

\begin{document}
\title{The roles of nucleon resonances in $\Lambda$(1520)
photoproduciton off proton}

\author{Jun He}
\email{junhe@impcas.ac.cn}
\author{Xu-Rong Chen}
\affiliation{
Institute of Modern Physics, Chinese Academy of Sciences, Lanzhou
730000, China}
\affiliation{
Research Center for Hadron and CSR Physics, Institute of Modern Physics of CAS
and Lanzhou University, Lanzhou 730000, China}
\date{\today}

\begin{abstract}

In this work the roles of the nucleon resonances in the
$\Lambda$(1520) photoproduction off proton target are investigated
within the effective Lagrangian method. Besides the Born terms,
including contact term, $s$-, $u$- and $K$  exchanged $t$-channels,
vector meson $K^*$ exchanged $t-$channel is considered in our
investigation, which is negligible at low energy and important at high
energy. The important nucleon resonances predicted by the constituent
quark model (CQM) are considered and the results are found well
comparable with the experimental data. Besides the dominant
$D_{13}(2080)$, the resonance $[\textstyle{5\over
2}^-]_2(2080)$ predicted by the
CQM is found important to reproduce the experimental data. Other nucleon
resonances are found to give small contributions in the channel
considered in this work. With all important nucleon resonances
predicted by CQM, the prediction of the differential  cross section at
the energy up to $5.5$GeV are presented also, which can be checked by
the future CLAS experimental data.

\end{abstract}

\pacs{14.20.GK, 13.75.Cs, 13.60.Rj} \maketitle
\maketitle
\section{Introduction}

The study of nucleon resonance is an interesting area in hadron
physics.  As of now, the nucleon resonances
around 2.1~GeV are still in confusion. The CQM predicted about two
dozens of the nucleon resonances in this region, most of which are in
$n=3$ shell. Only a few of them have been observed as shown in PDG
with large uncertainties~\cite{PDG}. Among these nucleon resonances,
$D_{13}(2080)$ attracts much attentions due to its importance found
in many channels, such as $\gamma p\to K^*\Lambda$~\cite{Kim:2011rm}
and $\phi$ potoproduction ~\cite{Kiswandhi:2011cq}. In the analyses of
the data for $\gamma p\to \eta'p$ by Zhang $et\ al.$ and Nakayama $et\
al.$ the contribution of $D_{13}(2080)$ is also found important to
reproduce the experimental data~\cite{Zhang:1995uha,Nakayama:2005ts}.
However, in the recent work by Zhong $et\ al.$~\cite{Zhong:2011ht},
the bump-like structure around $W=2.1$~GeV is from the contribution of
a $n=3$ shell resonance $D_{15}(2080)$ instead of $D_{13}(2080)$. In
Ref.~\cite{He:2008uf}, the $\eta$ photoproduction off the proton is
studied in a chiral quark model, a $D_{15}$ resonance instead of
$D_{13}$ state with mass about 2090~MeV is also suggested to reproduce
the experimental data. Hence, more efforts should be paid to figure out
the nucleon resonances around 2.1~GeV.

After many years of efforts, a mount of experimental data for the
nucleon resonances below 2~GeV have been accumulated while these above
2~GeV are still scarce. Due to the high threshold of $\Lambda(1520)$
production, about 2~GeV, it is appropriate to enrich the acknowledge
of the nucleon resonances especially the ones with the mass larger
than 2~GeV.  There exist some old experiments for kaon photoproduciton
off the nucleon with $\Lambda$(1520). In the last seventies,
SLAC~\cite{Boyarski:1970yc} used a 11 GeV photon beam on hydrogen to
study the inclusive reaction $\gamma p \to K^+Y$, where Y represents a
produced hyperon. The LAMP2 collaboration at Daresbury
\cite{Barber:1980zv} studied the exclusive reaction $\gamma p \to
K^+\Lambda^*$ with $\Lambda^*\to pK^-$ at photon energies ranging from
2.8 GeV to 4.8 GeV.  In the recent years, excited by the claimed
finding of pentaquark $\Theta$ with mass about 1.540~GeV reported by
LEPS collaboration in 2003~\cite{Nakano:2003qx}, many experiments are performed in
the $\Lambda(1520)$ energy region due to the close mass of
$\Lambda(1520)$ and $\Theta$.  Though the pentaquark is doubtable
based on the later more precise experiments, many experimental data about
$\Lambda(1520)$ photoproduction are accumulated, which provides
opportunity to understand the reaction mechanism of $\Lambda(1520)$
photoproduction off nucleon and the possible nucleon resonances in
this reaction.

The LEPS experiment (labeled as LEPS09 in this work) in the Spring-8 measured the $\Lambda$(1520)
photoproduction with liquid hydrogen and deuterium targets at photon
energies below 2.4~GeV~\cite{Muramatsu:2009zp}. A large asymmetry of
the production cross sections in the proton and neutron channels was
observed at backward K$^{+/0}$ angles. It supports the conclusion by
Nam $et\ al.$~\cite{Nam:2010au} that the contact term, with
$t-$channel $K$ exchange under gauge invariance, plays the most
important role in the $\Lambda(1520)$ photoproduciton and the
contribution from resonances $D_{13}(2080)$ is small and negligible.
Differential cross sections and photon-beam asymmetries for the
${\gamma} p$ $\rightarrow$ $K^{+}\Lambda$(1520) reaction have been
measured by LEPS Collaboration  (labeled as LEPS10) with linearly polarized photon beams at
energies from the threshold to 2.4 GeV at 0.6$<\cos\theta_{\rm
CM}^{K}<$1~\cite{Kohri:2009xe}.  A new bump structure was found at
$W\simeq 2.1$ GeV in the cross sections. Xie $et\ al.$ suggested the
bump structure could be reproduced by including the resonance
$D_{13}(2080)$~\cite{Xie:2010yk}. For the polarized symmetry, the
theoretical result by Xie $et\ al.$ seems to have an inverse sign
compared with the LEPS10 data. The model by Nam $et\ al.$ also give near
zero polarized asymmetry~\cite{Kohri:2009xe}.

In Refs.~\cite{Capstick:1992uc,Capstick:1998uh}, the electromagnetic
and strong decays were studied
in the CQM, with which the contributions of the nucleon resonances
in the phtoproductions can be calculated in Ref.~\cite{Oh:2007jd}. In this work we will
investigate the $\Lambda$(1520) photoproduction within the effective
Lagrangian method, with the nucleon resonances included according to
the CQM prediction, to find the roles of nucleon resonances,
especially the $D_{13}$ and $D_{15}$ states around 2.1~GeV, played in
the reaction mechanism.

This paper is organized as follows. After introduction, we will
present the effective Lagrangian used in this work and the amplitudes
based on the effective Lagrangian.  The differential cross section
at the low energy and the prediction at high energy will be given in
section~\ref{Sec: Results}. Finally the paper ends with a brief
summery.

\section{FORMALISM}

\begin{figure}[ht!]
\includegraphics[bb=130 590 470 720, scale=0.93,clip]{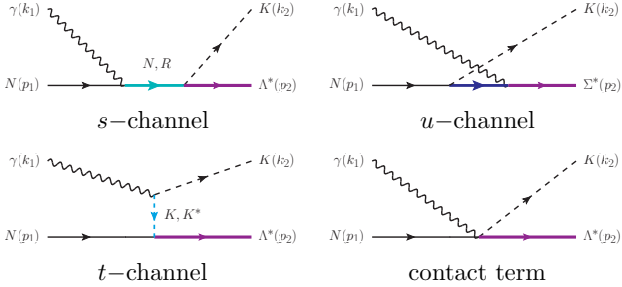}
\caption{(Color online) The diagrams for the $s$-, $u$-, $t$- channel
and contact term.}
\label{pic:dia}
\end{figure}

The mechanism for the $\Lambda(1520)$ photoproduction off nucleon with
$K$ is figured as the following diagrams in
Fig.~\ref{pic:dia}.
In this work, due to the small contribution from the $u$-channel as
shown in the literatures~\cite{Nam:2005uq,Nam:2010au,Xie:2010yk} it
is not considered in this work. Besides the Born terms,
including $s$-, $K$ exchanged $t$-channels and contact
term, the contributions from the $K^*$ exchanged $t$-
and resonance ($R$)
intermediate $s$-channels are considered in the current work.

\subsection{Born terms}

The Lagrangians used in the Born terms are given
below~\cite{Nam:2010au,Xie:2010yk},
\begin{eqnarray}
\label{eq:GROUND}
{\cal L}_{\gamma KK}&=&
ieQ_K\left[(\partial^{\mu}K^{\dagger})K-(\partial^{\mu}K)K^{\dagger}
\right]A_{\mu},
\nonumber\\
{\cal L}_{\gamma
NN}&=&-e\bar{N}\left[Q_N\rlap{/}{A}-\frac{\kappa_N}{4M_N}\sigma^{\mu\nu}
F^{\mu\nu}\right]N,
\nonumber\\
\mathcal{L}_{KN\Lambda^*}&=&\frac{g_{KN\Lambda^*}}{M_{\Lambda^*}}
\bar{\Lambda}^{*\mu}\partial_{\mu}K\gamma_5N\,+{\rm h.c.}, \cr
{\cal L}_{\gamma
KN\Lambda^*}&=&-\frac{ieQ_Ng_{KN\Lambda^*}}{M_{\Lambda^*}}
\bar{\Lambda}^{*\mu} A_{\mu}K\gamma_5N+{\rm h.c.},
\end{eqnarray}
where $F^{\mu\nu}=\partial^\mu A\nu-\partial^\nu A_\mu$ with $A^\mu$,
$N$, $K$, $\Lambda^*$ are the photon, nucleon, kaon and
$\Lambda^*(1520)$ fields. Here $Q_h$ is
the charge in the unit of $e=\sqrt{4\pi \alpha}$. The anomalous
magnetic momentum $\kappa=1.79$ for proton.  The
coupling constant $g_{KN\Lambda^*}=10.5$ obtained form the decay width of
$\Lambda^*\to NK$ in Particle Data Group~(PDG)~\cite{PDG}.

For $K^*$ exchange $t$-channel, we need the following Lagrangians,
\begin{eqnarray}
\mathcal{L}_{\gamma KK^{*}}&=& g_{\gamma
KK^{*}}\epsilon_{\mu\nu\sigma\rho}(\partial^{\mu}A^{\nu})
(\partial^{\sigma}K^{*\rho})K+\mathrm{h.c.}, \\
\mathcal{L}_{K^{*}N\Lambda^*}&=&
-\frac{ig_{K^*N\Lambda^*}}{m_{V}}\bar{\Lambda}^{*\mu}\gamma^{\nu}F_{\mu\nu}N
+\mathrm{h.c.},
\end{eqnarray}
with the coupling constant $g_{\gamma K^\pm K^{*\mp}}=0.254$~GeV$^{-1}$
extracted from decay
width in PDG~\cite{PDG}.  A Reggeized treatment will be applied to
$t$-channel. As discussion in
Ref~\cite{Toki:2007ab}, the coupling constant for the Reggeized treatment
can be different from the one in the effective Lagrangian approach.  Here we do
not adopt the value about 1 determined by $SU(6)$
model~\cite{Hyodo:2006uw}, but the larger one, about 10, by quark
model~\cite{Hyodo:2006uw} and phenomenological fitting
of Nam $et\ al.$~\cite{Nam:2005uq} and Totiv $et\ al.$~\cite{Titov:2005kf}.

The scattering amplitude for the photoproduction can be written
as follow:
\begin{eqnarray}
	-i{\cal T}_{\lambda_{\Lambda^*},\lambda_\gamma\lambda_N}
	=\bar{u}_\mu(p_2,\lambda_{\Lambda^*})A^{\mu\nu}u(p_1,\lambda_N)
	\epsilon_\nu(k_1,\lambda_\gamma)
\end{eqnarray}
where $\epsilon$, $u_{1}$, and ${u}^{\mu}_{2}$ denote the photon
polarization vector, nucleon spinor, and RS vector-spinor,
respectively. $\lambda_{\Lambda^*}$, $\lambda_\gamma$ and $\lambda_N$
are the helicities for $\Lambda(1520)$, photon and nucleon.

The amplitudes for Born $s$-,  $t$-channels and contact term are
\begin{eqnarray}
A^{\mu\nu}_{s}&=&-\frac{eg_{KN\Lambda^*}}{m_{K}}\frac{1}{s-M_N^2}
k_{2}^\mu{\gamma}_{5}\nonumber\\
&&\big[Q_N( (\rlap\slash p_1+M_N) F_{c}+\rlap{/}{k}_{1}F_{s})\gamma^\nu
\nonumber\\
&+&\frac{\kappa_{N}}{2M_{N}}
(\rlap{/}{k}_{1}+\rlap{/}{p}_{1}+M_{N})\gamma^\nu\rlap{/}{k}_{1} F_{s}
\big],
\\
A^{\mu\nu}_{t}&=&\frac{e Q_K g_{KN\Lambda^*}F_{c}}{m_K}\frac{1}{t-m_K^2}
q^{\mu}k_{2}^\nu\gamma_{5},
\\
A^{\mu\nu}_{\mathrm{cont.}}
&=&\frac{e Q_N g_{KN\Lambda^*}F_{c}}{m_K}
g^{\mu\nu}{\gamma}_{5},
\end{eqnarray}
where $s$, $t$, and $u$ indicate the Mandelstam variables.

The amplitude for $K^*$ exchange is
\begin{eqnarray}
A^{K^{*}}_{t}&=&
\frac{ig_{\gamma{K}K^*}g_{K^{*}NB}F_{V}}{m_{K^{*}}}
\frac{1}{t-m^{2}_{K^*}}
(\epsilon_{\sigma\rho\xi\nu}k^{\rho}_{1} k^{\xi}_{2})\nonumber\\
&&\gamma_\nu\left[(k_1^\mu-k_2^\mu)g^{\nu\sigma}+
(k_1^\nu-k_2^\nu)g^{\mu\sigma}
\right].
\label{amplitudes}
\end{eqnarray}

The form factors are introduced in the same form,
\begin{eqnarray}
F_i&=&\left(\frac{n\Lambda_i^4}{n\Lambda_i^4+(q_i^2-M_i^2)^2}\right)^n,\\
F_c&=&F_s+F_t-F_sF_t
\end{eqnarray}
where $i$ means $s$, $t$, $V$ and $R$ corresponding to $s$-,
$t$-channel, vector exchange $t$-channel and resonance intermediate
$s$-channel.

To describe the behaviour at the high photon energy,
we introduce the pseudoscalar and vector strange-meson Regge
trajectories~following~\cite{Guidal:1997hy,Corthals:2006nz,Titov:2005kf}:
\begin{eqnarray}
\label{eq:RT}
\frac{1}{t-m^{2}_{K}}\to\mathcal{D}_{K}
&=&\left(\frac{s}{s_{0}} \right)^{\alpha_{K}}
\frac{\pi\alpha'_{K}}{\Gamma(1+\alpha_{K})\sin(\pi\alpha_{K})},
\cr
\frac{1}{t-m^{2}_{K^{*}}}\to
\mathcal{D}_{K^{*}}&=&\left(\frac{s}{s_{0}} \right)^{\alpha_{K^{*}}-1}
\frac{\pi\alpha'_{K^*}}{\Gamma(\alpha_{K^*})\sin(\pi\alpha_{K^*})},
\end{eqnarray}
where $\alpha'_{K,K^{*}}$ indicates the slope of the trajectory.
$\alpha_{K, K^{*}}$ is the linear trajectory of the meson for even
or odd spin, which is a function of $t$ assigned as
follows,
\begin{eqnarray}
\label{eq:TR}
\alpha_{K}&=&0.70\,\mathrm{GeV}^{-2}(t-m^{2}_{K}),
\\
\alpha_{K^{*}}&=&1+0.85\,\mathrm{GeV}^{-2}(t-m^{2}_{K^{*}}).
\end{eqnarray}

To restore the gauge invariance, we redefine the relevant amplitudes
as follow~\cite{Nam:2010au},
\begin{eqnarray}
\label{eq:WT1}
&&i\mathcal{M}_{K}+i\mathcal{M}^{E}_{s}+i\mathcal{M}_{c}
\nonumber\\&\to&
i\mathcal{M}^{\mathrm{Regge}}_{K}+(i\mathcal{M}^{E}_{s}
+i\mathcal{M}_{c})(t-M^{2}_{K})\mathcal{D}_{K}
\end{eqnarray}

The Reggeized treatment should work completely at high photon energies
and interpolate smoothly to low photon energy. It have been
considered by Toki $et\ al.$~\cite{Toki:2007ab} and Nam $et\
al.$~\cite{Nam:2010au} by introducing a weighting function. Here we
adopt the treatment by Nam $et\ al.$,
\begin{eqnarray}
\label{eq:R}
&&F_{c,v}\to\bar{F}_{c,v}\equiv
\left[(t-m^{2}_{K,K^{*}})\mathcal{D}_{K,K^{*}}\right]
\mathcal{R}+F_{c,v}(1-\mathcal{R}),\,\,\,\,
\end{eqnarray}
where $\mathcal{R}=\mathcal{R}_{s}\mathcal{R}_{t}$ with
\begin{eqnarray}
\label{eq:RSRT}
&&\mathcal{R}_{s}=
\frac{1}{2}
\left[1+\tanh\left(\frac{s-s_{\mathrm{Regge}}}{s_{0}} \right)
	\right],\nonumber\\
&&\mathcal{R}_{t}=
1-\frac{1}{2}
\left[1+\tanh\left(\frac{|t|-t_{\mathrm{Regge}}}{t_{0}} \right) \right].
\end{eqnarray}

In this work the values of the four parameters for the reggeized
treatment are chosen as Nam. $et\ al.$ as presented in
Table~\ref{Tab: Regge}.
\begin{table}[h!]
\caption{parameters for the reggiezed treatment with unit GeV$^2$.}
\renewcommand\tabcolsep{0.5cm}
\renewcommand{\arraystretch}{1.3}
\begin{center}
\begin{tabular}{llllllll}  \toprule[1pt]
 $s_{Reg}=3$  & $t_{Reg}=0.1$
            & $s_0=1$     & $t_0=0.08$     \\\bottomrule[1pt]
\end{tabular}
\end{center}
\label{Tab: Regge}
\end{table}

\subsection{Nucleon resonances}
\label{Sec:R}

In Ref.~\cite{Xie:2010yk}, the $D_{13}(2080)$ is considered to
reproduce the bump structure near 2.1~GeV. In this work
all resonances predicted by CQM will be considered.  The Lagrangians for the
resonances with arbitrary half-integer spin
are~\cite{Chang:1967zzc,Rushbrooke:1966zz,Behrends:1957}
\begin{eqnarray}
	\mathcal{L}_{\gamma N R(\frac{1}{2}^{\pm})} &=&\frac{e f_2}{2M_N}
	\bar{N} \Gamma^{(\mp)}\sigma_{\mu\nu}F^{\mu\nu} R \,+{\rm h.c.}, \\
\mathcal{L}_{\gamma N R(J^{\pm})} &=&\frac{-i^{n}f_1}{(2m_N)^{n}} \bar{B}^*
~\gamma_\nu \partial_{\mu_2}...\partial_{\mu_{n}}
F_{\mu_1\nu}\Gamma^{\pm(-1)^{n+1}}R^{\mu_1\mu_2...\mu_{n}}\nonumber\\
&+&\frac{i^{n+1}f_2}{(2m_N)^{n+1}} \partial_{\nu}\bar{B}^*
~ \partial_{\mu_2}...\partial_{\mu_{n}}
F_{\mu_1\nu}\Gamma^{\pm(-1)^{n+1}}R^{\mu_1\mu_2...\mu_{n}}\nonumber\\
&+&{\rm h.c.},\label{Eq:Lg}
\end{eqnarray}
where $R_{\mu_1\ldots\mu_n}$ is the field for the
 resonance with spin $J=n+1/2$, and
\begin{eqnarray}
	\Gamma^{(\pm)}=(i\gamma_5,1)
\end{eqnarray}
for the different parity of resonance. The Lagrangians are also adopted
by the previous works on the nucleon resonances with
spins $3/2$ or $5/2$~\cite{Nam:2010au,Xie:2010yk,Oh:2007jd,Wu:2009md}.

The Lagrangian for the strong decay can be written as
\begin{eqnarray}
\mathcal{L}_{RK\Lambda^*} &=& \frac{ig_2}{2m_{K}}
\partial_\mu K\bar{\Lambda}^*_{\mu} \Gamma^{(\pm)}R,+{\rm h.c.}, \\
\mathcal{L}_{RK\Lambda^*}
&=&\frac{i^{2-n}g_1}{m_P^{n}} \bar{B}^*_{\mu_1}~\gamma_\nu\partial_\nu
\partial_{\mu_2}...\partial_{\mu_{n}} P\Gamma^{\pm(-1)^{n}}~R^{\mu_1\mu_2...\mu_{n}}\nonumber\\
&+&\frac{i^{1-n}g_2}{m_P^{n+1}} \bar{B}^*_{\alpha}~\partial_{\alpha}\partial_{\mu_1}
\partial_{\mu_2}...\partial_{\mu_{n}} P\Gamma^{\pm(-1)^{n}}~R^{\beta\mu_1\mu_2...\mu_{n}}\nonumber\\
&+&{\rm h.c.}.\label{Eq:Ls}
\end{eqnarray}
The corresponding propagator for the arbitrary half-inter spin can be
found in Appendix~\ref{Sec:propagator}.

In this work the coupling constants $f_1$, $f_2$, $g_1$ and $g_2$ will
be determined by the radiative and strong decays of the nucleon
resonances.  For a $j=\frac12$ resonance, the magnitudes of the
coupling constants $f_1$ and $h_1$ can be determined by the radiative
and strong decay widths. However, for nucleon resonances with high
spin $j \ge \frac32$, the decay widths are not enough to determine
coupling constants, $f_1$ and $f_2$ for radiative decay or $h_1$ and
$h_2$ for strong decay~\cite{Oh:2007jd}. Therefore, we need to know
the decay amplitudes to determine the coupling constants uniquely.

For the radiative decay, the helicity amplitudes are important
physical quantities and can be extracted from the experiment data of
photoproduction. The definition of the helicity amplitude is as below,
\begin{equation}
	A_\lambda=\frac{1}{\sqrt{2|\bm k|}}\langle \gamma({\bm
	k},1)~N (-{\bm k},\lambda-1)|-iH_\gamma|R({\bm
	0},\lambda)\rangle\label{Eq:helicity maplitudes}
\end{equation}
where $|\bm k|= (M_R^2 - M_N^2)/(2M_R)$, ${\bm k}$ is the momentum
of photon in the center of mass system of the decaying nucleon
resonance $R$, the $\lambda=1/2$ or $3/2$ is the helicity. Since there are two
amplitudes $A_{1/2}$ and $A_{3/2}$ for the resonances with $J>1/2$,
the coupling constants $f_1$ and $f_2$ in $\mathcal{L}_{RN\gamma}$
can be extracted from the helicity amplitudes of the resonance $R$.

The coupling constants $g_1$ and $g_2$ can be calculated analogously
by the following relation about the decay amplitude,
\begin{eqnarray}
&&\langle K(\bm{q})\, \Lambda^*(-\bm{q},m_f^{}) | -i\, \mathcal{H}_{\rm
int} | R (\bm{0}, m_J^{}) \rangle \nonumber\\&=& 4\pi M_R \sqrt{\frac{2}{q}}
\sum_{\ell,m_\ell^{}} \langle \ell\, m_\ell^{}\, \textstyle\frac32\,
m_f^{} | J \, m_J^{} \rangle\, Y_{\ell m_\ell^{}}(\hat{\bm q})
G(\ell),
\end{eqnarray}
where $\langle \ell\,m_\ell^{}\, \textstyle\frac32\, m_f^{} | J \, m_J^{} \rangle$
, $Y_{\ell m_\ell^{}}(\hat{\bm q})$ and $G(\ell)$are
Clebsch-Gordan coefficient, the spherical harmonics function and the partial wave decay
amplitude, respectively.
The explicit is presented in Appendix~\ref{Sec:cc}.

\section{Results}\label{Sec: Results}

With the Lagrangians presented in the previous section, the
$\Lambda(1520)$ photoproduction can be studied. With the
contributions from nucleon resonances determined by the decay
amplitudes, we will determine the parameters used in our
model first.  Then the observables, such as differential cross
section, will be calculated and compared with experiment.

\subsection{Contributions from nucleon resonances}

The different cross section have been measured by LEPS09 and LEPS10
experiments. As found by Nam
$et\ al.$ and Xie $et\ al.$~\cite{Nam:2010au,Xie:2010yk}, the most
important contribution at the low energy for the differential
cross section is from the contact term. The bump structure are from
$D_{13}(2080)$ suggested by Xie $et\ al.$ by fitting the LEPS10
data.  In this work we will use the helicity
amplitudes $A_\lambda$ and the partial wave decay amplitudes $G(\ell)$
to describe the decays of the resonances. In experiment, the helicity
amplitudes for the nucleon resonances with smaller mass are determined
well while the ones for the resonances with larger mass especially the
resonances with the mass larger than 2~GeV, which are considered in
this work, are still not well determined and can not be compared well
with the theoretical predictions.

In this work we will adopt the values obtained in the typical CQM by
Capstcik and Reborts~\cite{Capstick:1992uc,Capstick:1998uh} as input.
In their works the partial wave decay amplitudes are also provided.
The nucleon resonances considered in the calculation are listed in
Table~\ref{Tab: Resonances}. The threshold of $\gamma p\to
K\Lambda(1520)$ is about 2.01~GeV, so only the nucleon resonances
above 2.01~GeV  are included in the calculation. In the works by
Capstcik~\cite{Capstick:1992uc,Capstick:1998uh}, about two dozens of
nucleon resonances with spins up to $15/2$ are calculated. In the
current work, to simplify the claculation only the nucleon
resonances with large radiative decay and strong decay to
$\Lambda(1520) K$ are used in the calculation of $\Lambda(1520)$
photoproduction, which is reasonable in physics also.

\begin{table*}[htbp!]
\renewcommand\tabcolsep{0.45cm}
\renewcommand{\arraystretch}{1.3}
\caption{The nucleon resonances considered. The mass $m_R$, helicity
	amplitudes $A_\lambda$ and partial wave decay amplitudes
	$G(\ell)$ are in the unit of MeV, $10^{-3}/\sqrt{\rm{GeV}}$ and
	$\sqrt{\rm{MeV}}$, respectively. The last column is for $\chi^2$ after
turning off the corresponding nucleon resonance with $\chi^2=1.38$ in
full model.}
\begin{tabular}{ll|r|rr|rrr|rrr} \toprule[1pt]
State & PDG&  $M_R$ &  $A^p_{1/2}$  &  $A^p_{3/2}$  &  $G(\ell_1)$ &
$G(\ell_2)$ & $\sqrt{\Gamma_{\Lambda(1520) K}}$ & $\chi^2$ \\\hline
 $[N\textstyle{1\over 2}^-]_3(1945)$ &$N(2090)S_{11}$* &2090& 12    &
 & 6.4 $^{+ 5.7}_{- 6.4}$ & & 6.4 $^{+ 5.7}_{- 6.4}$ & 1.89\\
 $[N\textstyle{3\over 2}^-]_3(1960)$ &$N(2080)D_{13}$**& 2150& 36  & -43 & $-2.6 ^{+ 2.6}_{-
  2.8}$ & $-0.2 ^{+ 0.2}_{- 1.3}$ & 2.6 $^{+ 2.9}_{- 2.6}$& 12.42 \\
 $[N\textstyle{5\over 2}^-]_2(2080)$ & &2080& -3  & -14  & $-4.7 ^{+
  4.7}_{- 1.2}$ & $-0.3 ^{+ 0.3}_{- 0.8}$ & 4.7 $^{+ 1.3}_{- 4.7}$
  &4.01\\
 $[N\textstyle{5\over 2}^-]_3(2095)$ &$N(2200)D_{15}$** &2200& -2  & -6  & $-2.4 ^{+ 2.4}_{-
  2.0}$ & $-0.1 ^{+ 0.1}_{- 0.3}$ & 2.4 $^{+ 2.0}_{- 2.4}$ &1.59\\
 $[N\textstyle{7\over 2}^-]_1(2090)$ &$N(2190)G_{17}$****&2190& -34  & 28  & $-0.5 ^{+ 0.4}_{- 0.6}$ & 0.0
  $^{+ 0.0}_{- 0.0}$ & 0.5 $^{+ 0.6}_{- 0.4}$ &1.48\\
  $[N\textstyle{7\over 2}^+]_2(2390)$ &&2390&  -14 & -11   & 3.1 $^{+ 0.8}_{- 1.2}$ & 0.3 $^{+
  0.3}_{- 0.2}$ & 3.1 $^{+ 0.8}_{- 1.2}$ &1.79 \\
\bottomrule[1pt]
\end{tabular}
\label{Tab: Resonances}
\end{table*}

For the masses of the nucleon resonances, the suggested values
provided by PDG~\cite{PDG} are preferred and for the resonances not
listed in PDG, the prediction by CQM will be adopted.  Due to the
dominance of $D_{13}(2080)$ in the current channel, we adopt the mass
suggested by Xie $et\ al.$~\cite{Xie:2010yk}, 2150~MeV, which is a
little larger than the suggested value by PDG, 2080~MeV. In order to
prevent the proliferation of the free parameters, the decay widths for
all nucleon resonances are set to 200~MeV. Regarding the cut-offs for the
nucleon resonances, the typical value $\Lambda_R=1$~GeV with $n=1$ is used.
Therefore, the contributions from the nucleon resonances are fixed
because the helicity amplitudes and partial wave decay amplitudes
predicted by the CQM are adopted in the calculation to determine the
coupling constants.

\subsection{Determination of model parameters}

The contributions from the Born terms are very important as shown in
the previous works. The magnitudes of contributions from contact term, $u-$,
$s-$ and $K$ exchanges $t-$ channels are only depended on the coupling
constant $g_{KN\Lambda^*}$, which is determined by experimental  decay
width of $\Lambda^*\to NK$, and the cut off $\Lambda$. To reproduce
the LEPS10 data, a cut-off $\Lambda=0.6$~GeV and $n=1$ should be used. With this
setting, the $s-$ channel contribution can be found very small.

\begin{figure}[h!]
  \includegraphics[ bb=40 300 305 515,scale=0.85,clip]{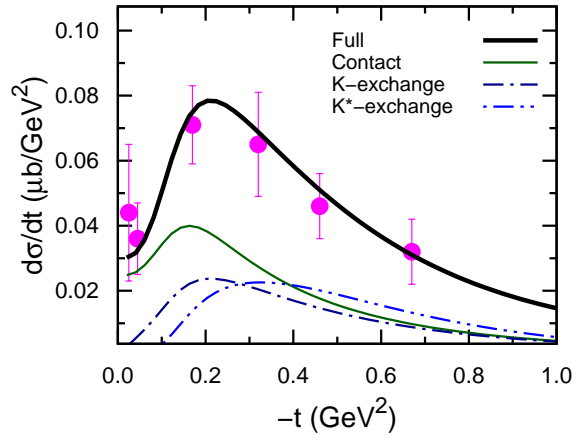}
  \caption{(Color online) The differential cross section $d\sigma/dt$ at 11~GeV and compared with the
  data by SLAC~\cite{Boyarski:1970yc}. The thick full line is for the result in full
  model. The full, Dash-dotted and Dash-dot-dotted lines are for the
  contact term, $K$-exchange channel and $K^*$-exchange channel. } \label{Fig: SLAC71DSDT}
\end{figure}

Now we turn to the contribution of vector meson $K^*$-exchange. We
consider  the differential cross section at 11~GeV by SLAC at
11~GeV, where the contributions from contact term and $t-$ channel
should be dominant, which is confirmed by the results shown in
Fig.~\ref{Fig: SLAC71DSDT}.

One can find at 11~GeV, the contribution from contact terms is still
most important while the contribution from the $K$-exchanged $t-$
channel becomes more important compared with that at the low energy.
However, it is still not enough to reproduce the experimental data by
including only the contributions from contact term and $K$-exchanged
$t-$ channel.  In the work by Nam $et\ al.$~\cite{Nam:2010au} the
differential cross section can be reproduced due to a larger cut-off
$\Lambda=0.675$~GeV adopted. However, with such value of $\Lambda$,
the differential cross section at low energy can not be well
reproduced. The parameters for the Reggeized
treatment, $s_{0,Reg}$, $t_{0,Reg}$ in Table~\ref{Tab: Regge}, are
varied and it is found impossible to compensate the deficiency. Here
we include the contribution of vector meson $K^*$ exchange. With
cutoff $\Lambda_V=0.8$~GeV and $n=2$, the differential cross section
by the SLAC experiment is well reproduced as shown in Fig.~\ref{Fig:
SLAC71DSDT}. In the figure one can find the contribution from vector
meson $K^*-$ exchange is comparable with the one from pseudoscalar
meson $K-$ exchange.

The determined cutoffs and the coupling constants for Born terms and
the $K^*$-exchange are collected in Table~\ref{cutoff}.

\begin{table}[h!]
\renewcommand\tabcolsep{0.35cm}
\begin{center}
\caption{The coupling constants for the $K$ and $K^*$ exchanged
channel and cut-offs in the unit of GeV.\label{cutoff}}
\begin{tabular}{lrlrlr}  \toprule[1pt]
 $g_{\gamma K^\pm K^{*\mp}}$&$0.254$ & $g_{KN\Lambda^*}$&$10.5$  &$\bar{g}_{K^*N\Lambda^*}$ & $10$\\
  $\Lambda$ & $0.6$    & $\Lambda_V $ & $0.8$ &$\Lambda_R$ &1
	    \\\bottomrule[1pt]
\end{tabular}
\end{center}
\end{table}

\subsection{Differential cross section}
\begin{figure*}[hbtp!]
  \includegraphics[ bb=130 395 550 720,scale=1.,clip]{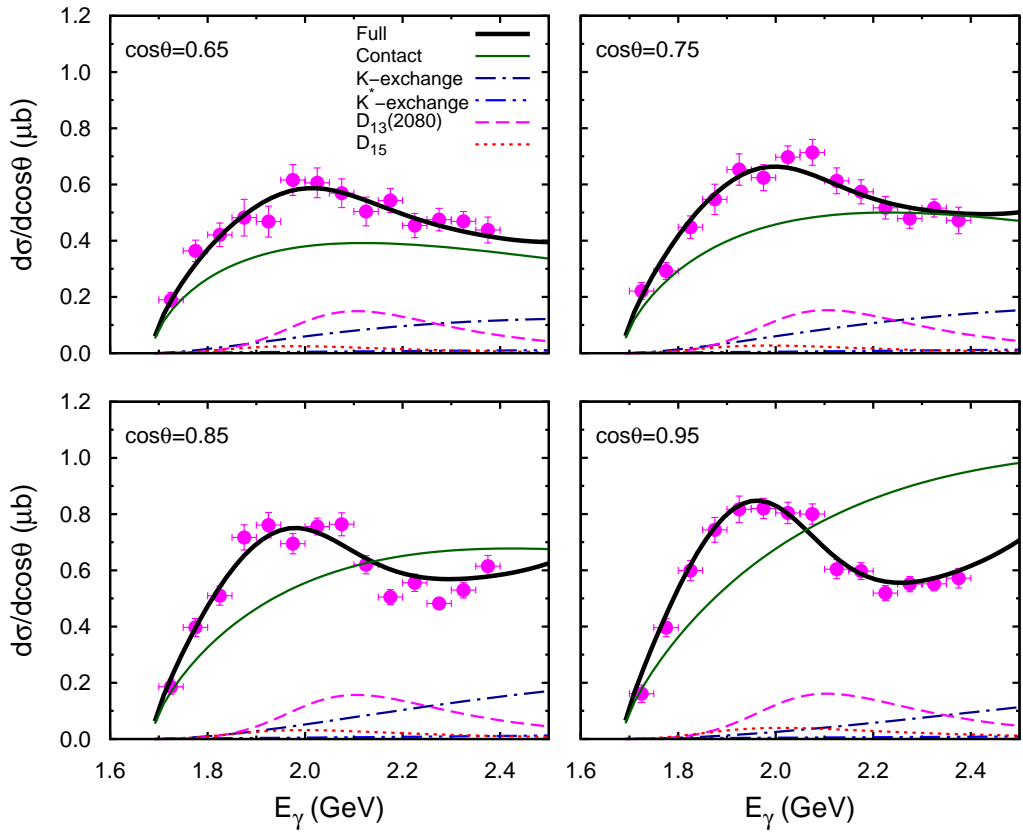}
  \caption{(Color online) The differential cross section  $d\sigma/d\cos\theta$
  at low photon energy and compared with the
  data by LEPS10. The thick full line is for the results in
  full model.The full, Dash-dotted and Dash-dot-dotted lines are for
  contact term, $K$-exchange channel and $K^*$-exchange channel. The dashed (dotted) line is
  for nucleon resonance $D_{13}(2080)$ ($D_{15}$).}\label{Fig: LEPS10dcs}
\end{figure*}

With the parameters determined and the amplitudes presented in
Table~\ref{Tab: Resonances}, the theoretical results of differential
cross section at the low energy
are presented in Fig.~\ref{Fig: LEPS10dcs} and compared with
the LEPS10 experiment.
As shown in the figure, the experimental data are well
reproduced in our model. The dominant contributions are from Born terms where the contact term play most
important role. At the low energy the $K$ exchange contribution
is smaller but visible while the vector meson $K^*$ exchanged contribution
is negligible.

To find the importance of each nucleon resonance, the $\chi^2$ for
the differential cross section from threshold to 2.5~GeV by LEPS10
experiment and these at 11~GeV by SLAC is calculated. $\chi^2$
obtained in the current work is 1.38, which is a little larger than
the 1.2 obtained by Xie $et\ al.$ but close to 1.4 in their fitting
with strong coupling constant as free parameters~\cite{Xie:2010yk}.
We would like to remind that in the current work both electromagnetic
and strong coupling constants are determined from CQM predictions with
the physics choice of the mass and cut-off for the nucleon resonance.

To check the role of nucleon resonance played in $\Lambda(1520)$
photoproduction, we list the $\chi^2$ after truing off the
corresponding nucleon resonance in Table~\ref{Tab:
Resonances}.  Compared with the value $\chi^2=1.38$ in the full model,
$\chi^2$ after turning off the $D_{13}(2080)$ increases significantly
to 12.42, which confirms the dominant role played by this resonance as suggested by Xie $et\ al.$~\cite{Xie:2010yk}. Besides
the dominant $D_{13}(2080)$, the contribution from a $D_{15}$
resonances is also important. After turning off $[\textstyle{5\over
2}^-]_2(2080)$, the $\chi^2$ will increase to 4.01 due to the
interference effect mainly. Simultaneously the $[\textstyle{5\over
2}^-]_3(2095)$ is found not important with $\chi^2=1.59$ after being
turned off.  In PDG~\cite{PDG} a $D_{15}$ state $N(2200)$ is listed , which is
assigned to the $[\textstyle{5\over 2}^-]_3(2095)$ in the CQM
usually~\cite{Capstick:1998uh}. The predicted decay amplitudes in
$N\pi$ and $\Lambda K$ channels for $[\textstyle{5\over 2}^-]_3(2095)$
are close to those for $[\textstyle{5\over
2}^-]_2(2080)$~\cite{Capstick:1998uh}. To check whether the $D_{15}$
state $[\textstyle{5\over 2}^-]_2(2080)$ instead of
$[\textstyle{5\over 2}^-]_3(2095)$ is the observed resonance $N(2200)$
listed in PDG, we vary the mass of $[\textstyle{5\over 2}^-]_2(2080)$
to 2.2~GeV and a $\chi^2$ about 4 is found. It indict that the
experimental data require the mass of $[\textstyle{5\over
2}^-]_2(2080)$ should be about 2.08~GeV. Hence $[\textstyle{5\over
2}^-]_2(2080)$ should not be assigned to $N(2200)$ due to the large
mass discrepancy about 100~MeV.

\begin{figure}[htbp!]
\includegraphics[ bb=51  295 299 438,scale=0.93,clip]{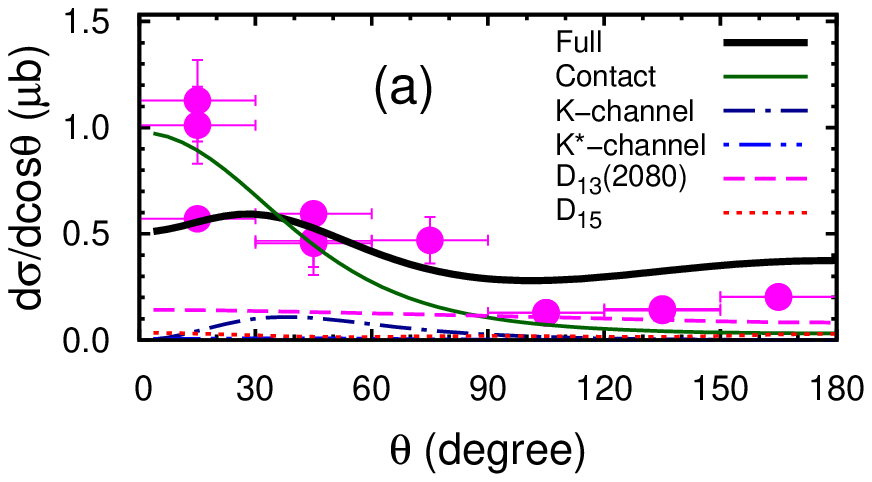}\\
\includegraphics[ bb=130  610 365 720,scale=1.0,clip]{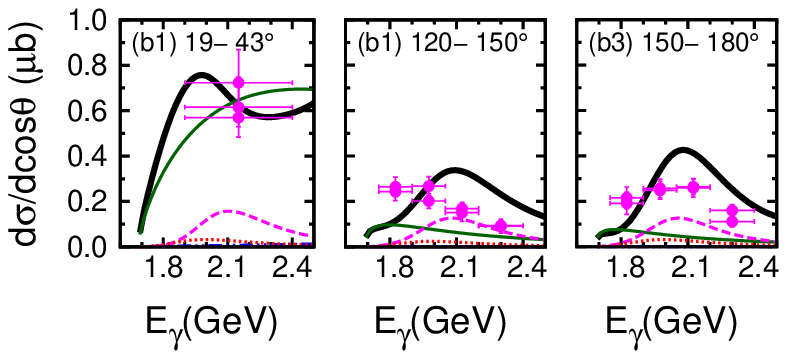}
\caption{(Color online) (a) The differential cross section
	$d\sigma/d\cos\theta$ with the variation of $\theta$ at
photon energy 2.2~GeV compared with the data by LEPS09 at photon
energy 1.9-2.4Ge. (b1-b3) The differential cross section
$d\sigma/d\cos\theta$ with the variation of photon energy $E_{\gamma}$
.
Notation as in Fig.~\ref{Fig: LEPS10dcs}.}\label{Fig: LEPS09dcs06}
\end{figure}

To give a picture around the resonance poles, we calculate the
$d\sigma/d\cos\theta$ aginst $\theta$ at 2.2~GeV and compared with LEPS09
data as shown in Fig.~\ref{Fig: LEPS09dcs06}.

The experimental data are reproduced generally in our model. One can
find the general shape for the differential cross section against
$\theta$ is mainly formed by the contact term contribution. The slow
increase in the backward is from the resonances contributions. For
differential cross section with the variation of $E_\gamma$, the
nucleon resonances  give contributions larger than the contact term
and are responsible to the bump at the backward angles.

\subsection{Polarized asymmetry}

\begin{figure}[h!]
  \includegraphics[ bb=50 300 290 510,scale=0.85,clip]{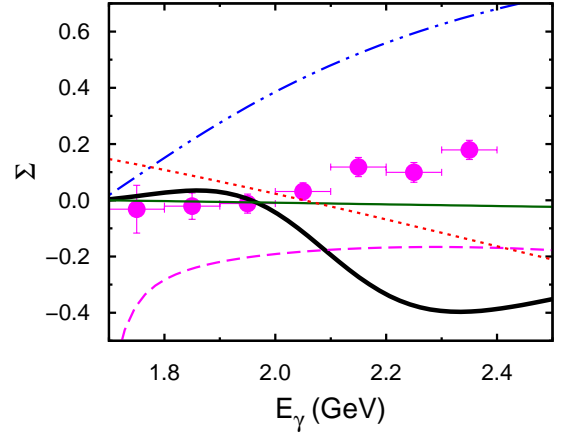}
\caption{(Color online) Polarization asymmetry and compared
with LEPS10. Notation as in Fig.~\ref{Fig: LEPS10dcs}}\label{Fig: LEPS10POL}
\end{figure}

The polarized asymmetry was measured in the LEPS10
experiment. Nearly zero polarized asymmetry was
obtained in the model by Nam $et\ al.$~\cite{Kohri:2009xe} . In
Ref.~\cite{Xie:2010yk} the result suggested the sign should be
reversed compared with the experiment. In this work, the similar
result as Ref~\cite{Xie:2010yk} is obtained as shown in
Fig.~\ref{Fig: LEPS10POL}. It suggests that further improvement
may be needed based on the effective Lagrangian method,
such as the lack of unitary, coupled-channel effects.

\subsection{Prediction of differential cross section at high energy}

\begin{figure}[h!]
  \includegraphics[ bb= 130 535 500 720 ,scale=0.71,clip]{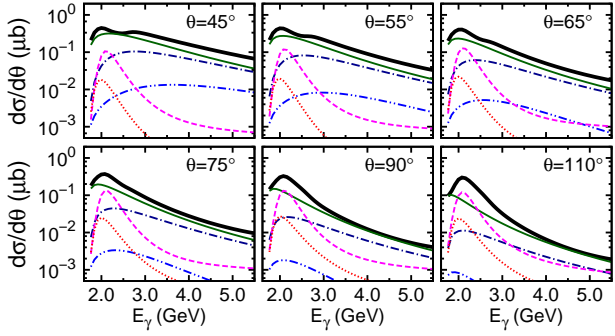}
  \caption{(Color online) The differential cross section $d\sigma/d\theta$ with the
	  variation of photon energy $E_\gamma$  predicted in this
	  work.
	  Notation as in Fig.~\ref{Fig: LEPS10dcs}}\label{Fig: CLAS10dcsp}
\end{figure}

In the above sections, all parameters, such as cut-offs, are
determined and the contributions from nucleon resonances are
introduced by the CQM predictions.  With these parameters the
differential cross sections at energy below 2.5GeV and at the energy
11~GeV have been reproduced well.  Hence it is possible to give the
prediction in the higher energy than 2.5~GeV. Recently CLAS
collaboration reported that  the $\Lambda(1520)$ photoproduction was
measured at the photon energy from $1.87$ to $5.5$~GeV and some
preliminary results have been obtained in the $eg3$
run~\cite{Zhao2011}. The $g_{11}$ experiment also run in this energy
region. Therefore, it is meaningful to make predictions.  The
differential cross section  $d\sigma/d\theta$ predicted by the model obtained in this
work are presented in Fig.~\ref{Fig: CLAS10dcsp}.

The contributions from $D_{13}(2080)$ and $D_{15}(2080)$ are still most
important as well as at the low energy. The $[\textstyle{7\over
2}^-]_2(2390)$ is small at high energy though it has higher
mass.
The contribution from the contact term plays most important role
at the energy up to 5.5~GeV especially at the forward angles
while the contributions from two resonances will decrease rapidly at
the photon energy large than the energy point corresponding to the
Breit-Wigner mass.

\begin{figure}[h!]
  \includegraphics[ bb= 130 420 400 720 ,scale=0.96,clip]{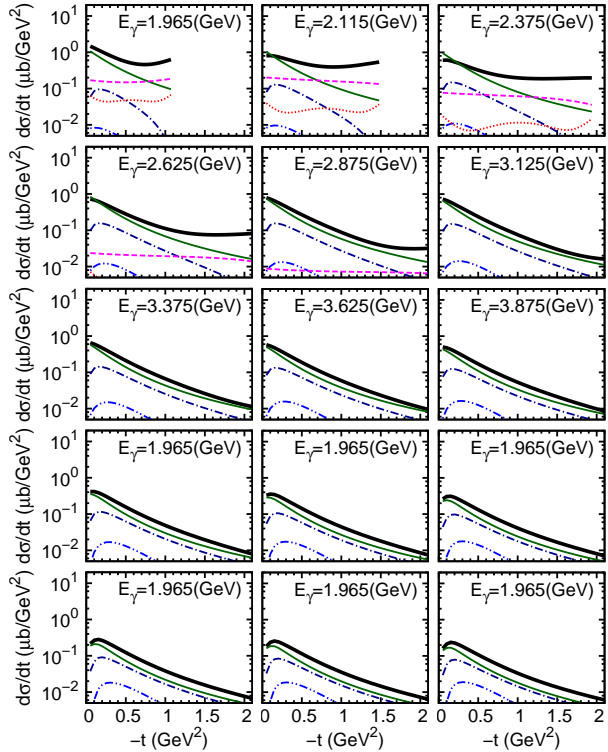}
  \caption{(Color online) The differential cross section $d\sigma/dt$ with variation
	  of $-t$ predicted in this work. Notation as in Fig.~\ref{Fig: LEPS10dcs}}\label{Fig: CLAS10DSDT}
\end{figure}

The dominance of contact contribution at the higher photon energy
can be observed more obviously in the differential cross section
$d\sigma/dt$ as shown in Fig.~\ref{Fig: CLAS10DSDT}.

One can find at high photon energy, the contact contribution is several times
larger than the vector meson $K^*$ exchange and the contributions from
the resonances are negligible. The contributions from the nucleon
resonances $D_{13}(2080)$ and $D_{15}$ are important at low energy
, and give a slow increase in the large $|t|$ region.

\subsection{Total cross section}

\begin{figure}[htbp!]
  \includegraphics[ bb=45 300 300 510 ,scale=0.85,clip]{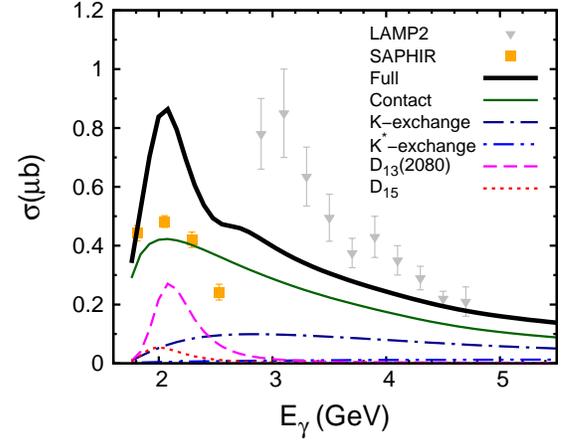}
  \caption{(Color online) Total cross section $\sigma$ with the
	  variation of the energy of photon $E_\gamma$.
  Notations for the theoretical results as in Fig.~\ref{Fig:
  LEPS10dcs}. The data are from
  Refs.~\cite{Barber:1980zv,Wieland:2010cq}.}\label{Fig: TCS}
\end{figure}

As of now, there only exist a few experimental data about the total
cross section.  The data by SAPHIR experiment
 are only at the low energy~\cite{Wieland:2010cq}
while LAMP2 experiment gives some data at high energy~\cite{Barber:1980zv}. In Fig.~\ref{Fig: TCS},
we show the theoretical results in our model and compared with the
experimental data.

As shown in the Fig.~\ref{Fig: TCS}, though two sets of data do
not have overlap in energy, it can be easily found that they are not
consistent with each other. The old LAMP2 data are higher than the
SAPHIR data systemically.  Our result is just in the middest
of two sets of data. As inidcted in the differential cross section, the
nucleon resonances $D_{13}(2080)$ and $D_{15}$ are responsible to the
peak around 2.1~GeV.

\section{Summery}

In this work we investigate the $\Lambda$(1520) photoproduction in the
$\gamma N \to K\Lambda(1520)$ reaction within the effective Lagrangian
method.  The contact term is dominant in the interaction mechanism and
$K$ exchanged $t-$channel are important except at the energy near the
threshold. The $K^*$-exchange $t$-channel plays important role in the high energy at 11~GeV
but is negligible at low energy.

The contributions of nucleon resonances are determined by the
radiative and strong decay amplitudes predicted by the constituent
quark model. The results shows that $D_{13}(2080)$ is the most
important nucleon resonance in $\Lambda(1520)$ photoproduction and
responsible to the bump structure in the LEPS10 experiment. A nucleon
resonance $[\textstyle{5\over 2}^-]_2(2080)$  predicted by CQM with
mass about $2100$~MeV, which can not befassigned as $N(2200)$, is also
essential to reproduce the experimental data around 2.1~GeV. The
contributions from other nucleon resonances are small even negligible.

With the contributions from Born terms and nucleon resonances, the
experimental data about differential cross sections can be reproduced
while there exists large discrepancy between experimental and theoretical
results in polarized asymmetry, which suggests further improvement,
such as including the coupled-channel effect,is required.
The predictions about the differential cross section at energy
1.75~GeV$<$E$_\gamma$$<$5.50~GeV are presented, which can be checked
by the future experimental data in the CLAS $eg3$ and $g11$ runs.

\section*{Acknowledgement}

This project is partly supported by the National Natural Science Foundation
of China under Grants No. 10905077, No.
11035006.

\appendix
\section{propagator}
\label{Sec:propagator}

In this appendix, we will present the propagator of half-integral spin
particle used in the current work, which is in the same theoretical
frame of the Lagrangians in Eqs.~(\ref{Eq:Lg},\ref{Eq:Ls})
\cite{Chang:1967zzc,Rushbrooke:1966zz,Behrends:1957}. The explicit
form of the propagator is 
\begin{eqnarray}
G^{n+\frac{1}{2}}_{R}&=&\frac{P^{n+\frac{1}{2}}_{\mu_{1}\mu_{2}...\mu_{n}\nu_{1}\nu_{2}...\nu_{n}}}
{p^{2}-m_{R}^{2}+im_{R}\Gamma_{R}},\\
P^{n+\frac{1}{2}}_{\mu_{1}\mu_{2}...\mu_{n}\nu_{1}\nu_{2}...\nu_{n}}
&=&\frac{n+1}{2n+3}
(\not\! p+ m)\gamma^{\alpha}\gamma^{\beta}P^{n+1}_{\alpha\mu_{1}\mu_{2}...\mu_{n}\beta\nu_{1}\nu_{2}...\nu_{n}},\
 \ \ \ \ \
\end{eqnarray}
where
\begin{eqnarray}
&&P^{n}_{\mu_{1}\mu_{2}...\mu_{n}\nu_{1}\nu_{2}...\nu_{n}}\nonumber\\&=&\left(\frac{1}{n!}\right)^2\sum_{P_{[\mu]}P_{[\nu]}}
\big[\prod_{i=1}^{n}
\tilde{g}_{\mu_{i}\nu_{i}}+a_{1}\tilde{g}_{\mu_{1}\mu_{2}}\tilde{g}_{\nu_{1}\nu_{2}}\prod_{i=3}^{n}
\tilde{g}_{\mu_{i}\nu_{i}}\nonumber\\
&&+...\nonumber\\
&&+a_{r}\tilde{g}_{\mu_{1}\mu_{2}}\tilde{g}_{\nu_{1}\nu_{2}}\tilde{g}_{\mu_{3}\mu_{4}}
\tilde{g}_{\nu_{3}\nu_{4}}...\tilde{g}_{\mu_{2r-1}\mu_{2r}}\tilde{g}_{\nu_{2r-1}\nu_{2r}}\prod_{i=2r+1}^{n}
\tilde{g}_{\mu_{i}\nu_{i}}\nonumber\\&&+...\nonumber\\
&&+\{^{a_{n/2}\tilde{g}_{\mu_{1}\mu_{2}}\tilde{g}_{\nu_{1}\nu_{2}}...\tilde{g}_{\mu_{n-1}\mu_{n}}\tilde{g}_{\nu_{n-1}\nu_{n}}(for\
even\
n)}_{a_{(n-1)/2}\tilde{g}_{\mu_{1}\mu_{2}}\tilde{g}_{\nu_{1}\nu_{2}}
...\tilde{g}_{\mu_{n-2}\mu_{n-1}}\tilde{g}_{\nu_{n-2}\nu_{n-1}}\tilde{g}_{\mu_n\nu_n}(for\
odd\ n)}\big], \\
&=&\left(\frac{1}{n!}\right)^2\sum_{P_{[\mu]}P_{[\nu]}}
\sum_{r=0}^{[n/2]}a_{r}\prod_{i=1}^{r}\tilde{g}_{\mu_{2i-1}\mu_{2i}}
\tilde{g}_{\nu_{2i-1}\nu_{2i}}\prod_{j=2r+1}^{n}
\tilde{g}_{\mu_{j}\nu_{j}},
\end{eqnarray}
with
\begin{eqnarray}
a_{r(n)}&=&\frac{\left(-\frac{1}{2}\right)^{r}n!}{r!(n-2r)!(2n-1)(2n-3)...(2n-2r+1)}.\label{bp2}\
\ \ \ \ \
\end{eqnarray}
where $\tilde{g}_{\mu\nu}=g_{\mu\nu}-\frac{q_{\mu}q_{\nu}}{q^{2}}$,
$P_{[\mu]}$ or $P_{[\nu]}$ means the permutations for $\mu$ or $\nu$,
and $[n]$ means integer round of $n$.

Some examples are presented in the  followings:
\begin{eqnarray}
G^{\frac{1}{2}}_{R}&=&\frac{(\not\! p + m)}{p^{2}-m_{R}^{2}+im_{R}\Gamma_{R}},\\
G^{\frac{3}{2}}_{R}&=&G^{\frac{1}{2}}_{R}
\left(-\tilde{g}_{\mu\nu}+\frac{1}{3}\tilde{\gamma}_{\mu}
\tilde{\gamma}_{\nu}\right),\\
G^{\frac{5}{2}}_{R}&=&G^{\frac{1}{2}}_{R}
\sum^2_{P_{[\mu]},P_{[\nu]}}
\Big[\frac{1}{4}\tilde{g}_{\mu\nu}\tilde{g}_{\mu\nu}
-\frac{1}{20}\tilde{g}_{\mu\mu}\tilde{g}_{\nu\nu}
-\frac{1}{10}\tilde{\gamma}_{\mu}\tilde{\gamma}_{\nu}
\tilde{g}_{\mu\nu}\Big],\ \ \ \ \
\end{eqnarray}
where
$\tilde{\gamma}_{\nu}=\gamma_{\nu}-\frac{p_{\nu}~\not\!p}{p^{2}}$.

\section{Extracting the coupling constants}
\label{Sec:cc}

As mentioned in section~\ref{Sec:R}, the
Lagrangians in Eqs.~(\ref{Eq:Lg},\ref{Eq:Ls}) show both radiative and strong
decay of the nucleon resonance with $J>1/2$ are described by two
coupling constants. In this appendix we will show how to extract
coupling constants from the helicity  amplitudes and partial decay
amplitudes.

The helicity amplitudes for the resonances with spin-parity $J^P$ can
be calculated from the definition in Eq.~(\ref{Eq:helicity maplitudes})
easily in the $c. m.$ frame
as
\begin{eqnarray}
A_{\half}(\textstyle\frac12^{P}) &=&\mathcal{P}\frac{ef_1}{2M_N}
\sqrt{\frac{k_\gamma M_R}{M_N}},
 \\
{A}_{\thalf}(\textstyle J^P)
&=&
\mathcal{P} F_\thalf ~k_\gamma^{n-1}
~\frac{e}{\sqrt{2}(2M_N)^n}
\sqrt{\frac{k_\gamma M_R}{M_N}}\nonumber\\
&\cdot&\left[f_1
+\left(\frac{\mathcal{P}}{M_R}\right)
\frac{f_2}{4M_N} M_R(M_R+\mathcal{P} M_N)\right]
,\\
{A}_{\half}(\textstyle J^P)
&=&
\mathcal{P} F_\half ~k_\gamma^{n-1}
~\frac{e}{\sqrt{6}(2M_N)^n}\sqrt{\frac{k_\gamma M_N}{M_R}}\nonumber\\
&\cdot&\left[f_1+
\frac{f_2}{4M_N^2} M_R(M_R+\mathcal{P} M_N)\right]
\label{eq:RNgamma}
\end{eqnarray}
where $F_r=\prod_{i=2}^n\langle 10,i-\half~r|i+\half~r\rangle$ and
$\mathcal{P}=P (-1)^{n}$ with $n=J-\half$. Here $M_N$ and $M_R$ are the masses of
nucleon and the nucleon resonance, and $k_\gamma$ are the energy of
the photon. The helicity amplitudes $A_\lambda$ can be obtained
by CQM or extracted from the experiment. Now the coupling constants for
the resonances with $J=\half$ can be obtained from $A_{1/2}$ directly
and the ones with $J>\half$ can be solved from the two equations about
$A_{1/2}$ and $A_{3/2}$.

For the nucleon resonaces decay to $K$ and $\Lambda^*(1520)$, we have
\begin{eqnarray}
{\cal A}(\textstyle J^P,r,\theta)
&=&\langle K(\bm{q})\, \Lambda^*(-\bm{q},r) | -i\, \mathcal{H}_{\rm
int} | R (\bm{0}, r) \rangle \nonumber\\
&=& 4\pi M_R \sqrt{\frac{2}{q}}
\sum_{\ell} \langle \ell\, 0\, \textstyle\frac32\,
r| J \, r \rangle\, Y_{\ell 0}G(\ell),
\end{eqnarray}
if we choose $m_\ell=0$. Here $\theta$ is the angle of the final $K$. The relative orbital angular momentum $\ell$ of
the final state is constrained by the spin-parity of the resonance.
For the nucleon resonances with $J>1/2$, there are two $\ell$ which is
marked as $\ell_1$ and $\ell_2$ thereafter.

Since partial wave decay amplitudes $G(\ell)$ is independent on the
$\theta$, we choose $\theta=0$ and reach 
\begin{eqnarray}
A(J^P,\theta=0,r)
&=&  \sqrt{\frac{2\ell_1+1}{4\pi}}
\langle \ell_1\, 0\, {\textstyle\frac32}\,
r| J \, r \rangle\, G(\ell_1)\nonumber\\
&+& \sqrt{\frac{2\ell_2+1}{4\pi}}\langle \ell_2\, 0\, \textstyle\frac32\,
r| J \, r \rangle\, G(\ell_2)
\end{eqnarray}
The $G(\ell)$ can be obtained from the CQM and is independent on $r$, so the difference of 
the amplitudes with different $r$ is from the Clebsch-Gordan coefficient.

With certain $J>\half$ and $r$, the decay amplitudes can be calculated and
has the form
\begin{eqnarray}
&&{A}(\textstyle J^P,r,\theta=0)\nonumber\\
&=&-\frac{\sqrt{8\pi M_R}}{M_{\Lambda^*}} \frac{q^{n-1}}{m_K^n}
\nonumber\\
&\cdot&\Bigg[h_1{\cal P} [ c_{l_0{\cal P}}p^0+(c_{1{\cal
P}}-c_{l_0{\cal P}})M_{\Lambda^*}]
\sqrt{p^0+{\cal P}M_{\Lambda^*}}(M_R-{\cal P}M_{\Lambda^*})\nonumber\\
&+& \frac{h_2}{m_K} M_Rq^2~c_{l_0{\cal P}}\sqrt{p^0+{\cal P}M_{\Lambda^*}},
~ \Bigg]
\end{eqnarray}
where $c_{1\pm ,~l_0\pm}=\prod_{i=2}^n\langle 1 0i-\half r|i+\half r\rangle
[1,~\delta_{l_00}](|\langle 1 l_0\half
\half|\thalf r\rangle|^2\pm|\langle 1 l_0\half-\half|\thalf
r\rangle|^2)$ and
$\mathcal{P}=P(-1)^{n+1}$. Here $M_{\Lambda^*}$ ans $m_K$ are masses
of $\lambda(1520)$ and $K$ meson, and $p^0$ are the energy of the
final $\Lambda(1520)$. Now the coupling constants $h_1$ and $h_2$
are related to the partial wave decay amplitudes $G(\ell)$. The $h_1$
and $h_2$ can be solved from above equations by choosing 
$r=1/2$ and $r=3/2$.

\end{document}